\documentstyle[aps,prb,epsf,epsfig,twocolumn,floats]{revtex}

\begin{document}

\draft

\twocolumn[\hsize\textwidth\columnwidth\hsize\csname @twocolumnfalse\endcsname

\title{Charge-density waves in one-dimensional Hubbard
superlattices}
\author{Thereza Paiva and Raimundo R.\ dos Santos}
\address{Instituto de F\' \i sica, Universidade Federal do Rio de Janeiro,
                 Cx.P.\ 68.528, 21945-970 Rio de Janeiro RJ, Brazil\\}
                 
\date{Version 2.34 - \today}

\maketitle

\begin{abstract}
We study the formation of charge density waves (CDW's) in one-dimensional
Hubbard
superlattices, modeled by a repeated pattern of repulsive ($U>0$) and
free ($U=0$) sites. 
By means of  Lanczos diagonalizations for the ground state, we calculate
the charge structure factor. Our results  show that while the superlattice
structure affects the modulation of the charge density waves,  
the periodicity can still be predicted through an effective density. 
We also show that, for a fixed repulsive layer thickness,  the periodicity
of the CDW is an oscillatory function of the free layer thickness.
\end{abstract}

\pacs{PACS:     
      71.45.Lr, 
      73.21.Cd, 
      71.27.+a, 
      71.10.-w, 
      73.20.Mf, 
      72.15.Nj. 
}
\vskip2pc]

Charge-density waves (CDW's) are present in a variety of homogeneous
strongly correlated electron systems, especially some
quasi-one-dimensional organic
conductors.\cite{JS82,Pouget89,Jerome94,Gruner94,Brown94} An interesting
property of CDW systems is their non-ohmic behavior: CDW's incommensurate
with the underlying lattice are pinned by impurities, so that only
voltages above a certain threshold are able to produce observable
currents.  Just above this critical voltage, the current rises sharply,
indicating a huge increase in the conductivity.\cite{Gruner94,Brown94}

Thus, one of the key issues in the study of CDW's is to determine the
period of charge modulation in the ground state. From the theoretical
point of view, this has been addressed with the aid of various models, but
care must be taken even in the simplest case of the one-dimensional
Hubbard model. Indeed, according to the Luttinger liquid (LL)
description,\cite{Solyom79,Schulz90,Frahm90,Voit94} the large-distance
behavior of the charge density correlation function in the ground state
is given by
\begin{equation} 
\langle n(x)n(0)\rangle = {K_\rho\over (\pi x)^2}
           + A_1 { \cos (2 k_F x) \over x^{1+K_\rho} \ln^{3/2} x}\\
           + A_2 { \cos (4 k_F x) \over x^{4K_\rho}}, \label{nn}
\end{equation} 
where the amplitudes $A_1$ and $A_2$, and the exponent
$K_\rho$ are interaction- and density-dependent parameters; for the
Hubbard model with repulsive interactions one has\cite{Schulz90} ${1\over
2} \leq K_\rho < 1$, so that charge correlations are expected to be
dominated by the $2k_F$ term, where $2k_F=\pi\rho$ is the Fermi wave
vector for a density $\rho<1$ of free electrons on a periodic
lattice; if $\rho>1$, then $2k_F=\pi(2-\rho)$. By
contrast, numerical analyses predict that CDW correlations are dominated
by the $4k_F$ term, for strong enough on-site repulsion
$U$.\cite{HS,tclp00a} These two scenarios can only be reconciled if $A_1$
vanishes above some $U^*(\rho)$.\cite{tclp00a}

On the other hand, the development of diverse and very accurate deposition
techniques over the last two decades has generated a whole new class of
materials, generically called heterostructures. Systems made up of very
thin layers -- in some cases a few atoms thick -- of materials with
different properties have been grown; if the layered structure is
periodic, it is referred to as a superlattice (SL).  Interesting new phenomena
have emerged from these heterostructures, demanding novel theoretical
mechanisms to explain the observed data, and giving rise to improved
devices for technological applications.  

We have recently considered a model for strongly correlated electrons in
one-dimensional superlattices,\cite{tclp96,tclp98,tclp00b} defined
by the Hamiltonian
\begin{equation} 
\label{Ham} 
{\cal H}=-t\sum_{i,\ \sigma}
\left(c_{i\sigma}^{^{\dagger}} c_{i+1\sigma}+\text{H.c.}\right)  + \sum_i
U_i\ n_{i\uparrow}n_{i\downarrow} 
\end{equation} 
where, in standard notation, $i$ runs over the sites of a one-dimensional
lattice, $c_{i\sigma}^{^\dagger}$ ($c_{i\sigma}$) creates (annihilates) a
fermion at site $i$ in the spin state $\sigma=\uparrow\ {\rm or}\
\downarrow$, and $n_i=n_{i\uparrow}+n_{i\downarrow}$, with
$n_{i\sigma}=c_{i\sigma}^{^\dagger}c_{i\sigma}$; the on-site Coulomb
repulsion is taken to be site-dependent: $U_i=U>0$, for sites within the
repulsive layers, and $U_i=0$ otherwise.  The repulsive and free layers
have $L_U$ and $L_0$ sites, respectively; the `aspect ratio' is defined
as $\ell=L_U/L_0$. The extension of this model to a Luttinger liquid
superlattice (LLSL) has been discussed recently.\cite{llsl,llsl2} 

The magnetic properties of model (\ref{Ham}) turned out to be quite
different from those of the corresponding homogeneous
system;\cite{tclp96,tclp00b} also, the electronic density at which a
charge gap opens in these SL's can be fine-tuned by an appropriate choice
of $\ell$.\cite{tclp98,llsl,llsl2} In view of this, the behavior of CDW's
on a
superlattice structure should also be of interest. In particular, a
crucial question is how to identify and parametrize the dominant
charge-density correlations in the case of these Hubbard superlattices
(HSL's). Our purpose here is to address these points.

We use the Lanczos algorithm\cite{Roomany80,Gagliano86,Dagotto94} to
determine the ground state of Eq.\ (\ref{Ham}), for finite lattices of
$N_s$ sites with periodic boundary conditions, in the subspace of fixed
particle-density (canonical ensemble) $\rho=N_e/N_s$, where $N_e$
is the total number of electrons. The signature of a CDW instability
is a {\em cusp} at $q=q^*,$ in the charge-density structure factor,
\begin{equation}
C(q)={1\over N_c} \sum_{i,j} {\rm e}^{iq(r_i-r_j)}
\langle  Q_i Q_j \rangle\;,
\label{Cq}
\end{equation}
with
$\langle  Q_i Q_j \rangle= \langle \psi_0|  n_i n_j
|\psi_0\rangle-\langle
\psi_0 | n_i
| \psi_0 \rangle \langle \psi_0| n_j | \psi_0 \rangle$,
where  $| \psi_0 \rangle $ is the
ground state and  $N_c$ is
the number of periodic cells, $N_c=N_s/N_b$, for a basis with
$N_b=L_U+L_0$ sites.

\begin{figure} 
\begin{center} 
\epsfxsize=8.5cm
\epsffile{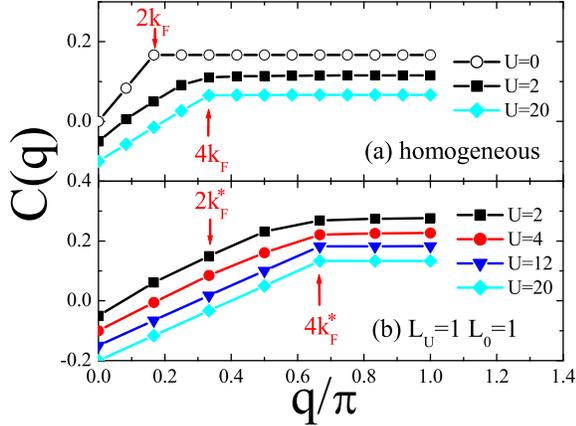} 
\caption{Charge structure factor for $\rho=11/6$ and $N_s=24$, for a
homogeneous chain (a), and for a superlattice with $L_U=L_0=1$
(b). The curves are labelled by the values of the on-site coupling:  
$U=0$ (open circles), $U=2$ (squares), $U=4$ (circles), $U=12$ (down
triangles), and $U=20$ (diamonds). Successive vertical shifts of 0.05 have
been imposed on the curves for clarity.}
\label{l11hom} 
\end{center}
\end{figure}

In order to study the behavior of $C(q)$ in the widest possible range of
values of $L_0$, we focus our discussion on the case of 24-site lattices,
with electron density $\rho=11/6$. Figure \ref{l11hom} compares the charge
structure factor for the homogeneous chain with that for a SL with $L_U=1$
and $L_0=1$. From Fig.\ \ref{l11hom}(a) one sees that the $2k_F$ cusp in
the free ($U=0$) homogeneous case is displaced to $4k_F$ as $U$
increases.\cite{tclp00a} The SL structure changes the periodicity of the
CDW as shown in Fig.\ \ref{l11hom}(b): the cusp is displaced relative to
the position for the homogeneous chain. Similarly to the homogeneous
system, the cusps sharpen as $U$ increases; their positions, however,
remain locked at $2\pi/3$ for that SL configuration and electronic
density.

\begin{figure} 
\begin{center} 
\epsfxsize=8.5cm
\epsffile{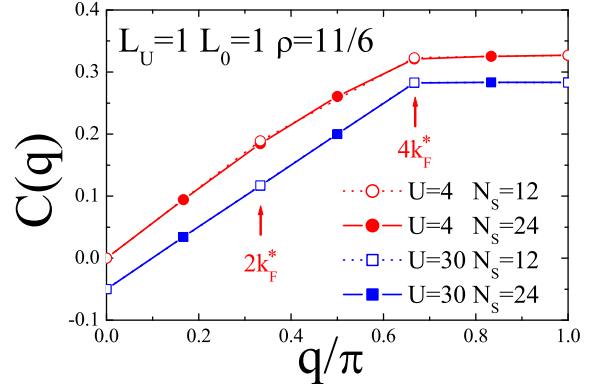} 
\caption{Charge structure factor for a superlattice with $L_U=L_0=1$,
$\rho=11/6$, and both $U=4$ (circles) and $U=30$ (squares). Data
corresponding to $N_s=24$ ($N_s=12$) are denoted by filled (open) symbols,
joined by full (dotted) lines. For clarity, successive vertical shifts of
0.05 have been imposed on the curves corresponding to different values of
$U$.}
\label{fse} 
\end{center}
\end{figure}

Before discussing the location of the cusps, let us check on the role of
finite-size effects. Recall that the appropriate finite-size parameter is
the number of repeating units, $N_c$.  Figure \ref{fse} shows $C(q)$ for
the above mentioned SL and for two different lattice sizes, namely
$N_s=12\ (N_c=6)$ and $N_s=24\ (N_c=12)$. One sees that for $U=4$ the cusp
is sharpened as one goes from $N_c=6$ to $N_c=12$; the cusp positions,
however, do not change as the size increases.  For large couplings ($U=30$
in the figure), the curves lie on top of each other, so that finite-size
roundings are not noticeable. Thus we can rule out finite size effects as
playing any crucial role in determining the cusp positions.\cite{tclp00a}

\begin{figure}[h] 
\begin{center} 
\epsfxsize=8.5cm
\epsffile{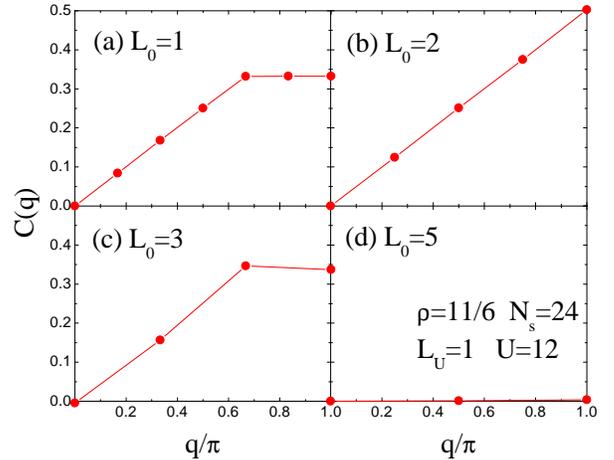} 
\caption{Charge structure factor for $\rho=11/6$, $U=12$, $N_s=24$,
$L_U=1$, and for different free layer lengths: (a) $L_0=1$; (b) $L_0=2$;
(c) $L_0=3$; (d) $L_0=5$.}
\label{cqsl} 
\end{center}
\end{figure}

The main feature determining the periodicity of the CDW can be identified
through a systematic study of $q^\ast$ as a function of layer thickness.
Figure \ref{cqsl} shows the charge structure factor for a fixed length of
the repulsive layer ($L_U=1$), and for different lengths of the free
layer: $L_0=1$, 2, 3, and 5. We see that the CDW modulation depends on the
superlattice: $q^*/\pi=2/3,$ 1, and 2/3 for $L_0=1,$ 2, and 3,
respectively; the vanishing $C(q)$ obtained for $L_0=5$ is associated with
$q^*=0$.

These cusp positions can be analysed in a strong coupling scenario. We
first recall that there is a special density, $\rho_I$, corresponding
to each free site being doubly occupied, while each repulsive site is 
singly occupied; it can be written as\cite{tclp98}
\begin{equation}
\rho_I = {2+\ell\over 1+\ell}.
\label{rhoI}
\end{equation}
At this density, a Mott-Hubbard gap opens and the system is an
insulator;\cite{tclp98,llsl,llsl2} that is, the transport properties of a
SL at
$\rho_I$ should be similar to that for a homogeneous chain at
half-filling,\cite{tclp98} which includes the breakdown of
CDW's.\cite{essler} This latter point is nicely illustrated by the
vanishing of $C(q)$ for all $q$ for the SL with $L_U=1$ and $L_0=5$, as
shown in Fig.\ \ref{cqsl}(d): $\rho_I$ is exactly 11/6 for that SL.
For densities above $\rho_I$, all free sites remain
doubly occupied, while the amount of charge on the repulsive sites can
fluctuate.\cite{tclp00b} Charge correlations between different cells are
therefore dominated by the way in which electrons are distributed within
each repulsive layer. Accordingly, we define the effective {\em cell
density} (i.e., the number of active electrons per unit cell),
\begin{equation}
\rho_{\rm eff}=\rho(L_0+L_U)-2L_0,
\label{rhoeff}
\end{equation}
where $\rho$ is the overall density. If one similarly defines an
effective Fermi wave vector, $k_F^*$, from
\begin{equation}
2k_F^*=\pi (2- \rho_{\rm eff}),
\label{kfeff}
\end{equation}
then the cusp in $C(q)$ is located simply at $q^*=4k_F^*$, in a way
completely analogous to that of the homogeneous system.  In Fig.\
\ref{l11hom}(b), $4k_F^\ast$ thus calculated is indicated by an
arrow, and it is clear that it coincides with the cusp positions.  By
the same token, for
all SL's shown in Fig.\ \ref{cqsl}, the position of the cusps also
perfectly match the $4k_F^*$ as obtained from Eq.\ (\ref{kfeff}).

\begin{figure}[ht]
\begin{center}
\leavevmode
\epsfxsize=8.5cm
\epsffile{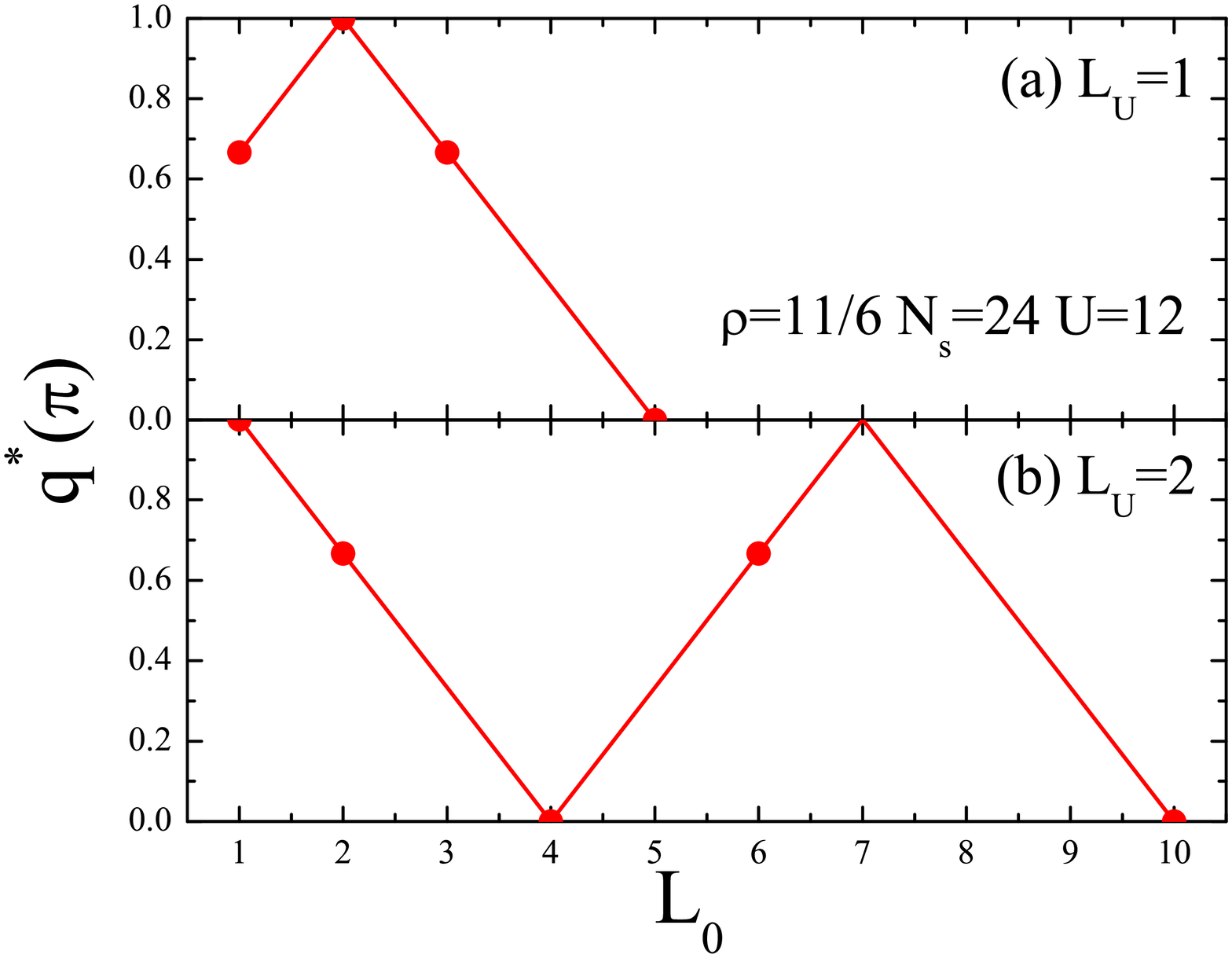}
\caption{$q^*$ as a function of $L_0$ for $N_S$=24, $U=12$, $\rho=11/6$,
and for different repulsive layer lenghts: (a) $L_U=1$, and (b) $L_U=2$.}
\label{qast}  
\end{center}
\end{figure}

It is now illustrative to discuss  how the periodicity of the CDW depends
on the width of the free layer as the repulsive layer width is kept fixed,
as done in Fig.\ \ref{qast} where $q^*$ is shown as a function of
$L_0$ for both $L_U=1$  and $L_U=2$. First, we see that the strong
coupling prediction, $q^\ast=4k_F$, with $k_F$ given by Eq.\
(\ref{kfeff}), is satisfied for any $U>0$: it agrees perfectly well with 
the data points (circles) obtained from Lanczos diagonalization of
different SL's with  $U=12$.
And, second, we see that $q^\ast$ is periodic with $L_0$, for a fixed
$L_U$. The period of oscillation
$\Delta L_0$ can be determined by imposing $q^*(L_0)=
q^*(L_0+\Delta L_0)$, which yields
\begin{equation}
\Delta L_0= (2-\rho)^{-1} = { \pi \over {2 k_F}}
\label{deltal0}
\end{equation}
where $\rho$ is the overall density. 

A similar oscillatory behavior has been found for the periodicity of SDW's
on Hubbard SL's,\cite{tclp00b} which was identified with the exchange
oscillation in magnetic multilayers; this latter feature is, in turn,
intimately connected with the Giant Magnetoresistance (GMR) effect (see
e.g., Ref.\ \onlinecite{pg01} and references therein). Thus, the above 
results indicate that one can also fine-tune a given charge distribution
by a suitable choice of the length of the spacer (i.e., free) material.

As a final remark, we should mention that experimental realizations of
(variations of) the model discussed here have recently appeared in the
literature.  Indeed, by growing $L_A$ layers of semimetallic (narrow gap
semiconductor) TiSe$_2$ in succession to $L_B$ layers of
metallic/superconducting NbSe$_2$ -- both compounds undergo CDW
transitions -- Noh {\it et al.}\cite{dcj00} were able to grow `superfilms'
1000 \AA\ thick, with $L_A$ and $L_B$ in the range 1 to 24 layers. The
superconducting critical temperature and the temperature dependence of the
conductivity were then obtained as functions of the SL structure.

In summary, we have considered the formation of CDW's in one-dimensional
Hubbard superlattices. We have established that the SL structure affects
the charge distribution in relation to the homogeneous lattices, though
one can still predict the new periodicity through an effective cell
density. It has also been established that the periodicity of the CDW is
an oscillatory function of the spacer thickness, for a fixed repulsive
layer thickness. As a consequence, for a given (overall) electronic
density, one can fine-tune a desired charge distribution by a suitable
choice of the length of the spacer (i.e., free) material. We hope the
present work stimulates further investigations on the superlattices made
up of transition metal dichalcogenides, focusing on the CDW periodicity
along the direction of growth.

\acknowledgments 
The authors are grateful to A.\ L.\ Malvezzi, E.\ Miranda,
and J.\ Silva-Valencia for useful discussions.  Financial support from the
Brazilian Agencies FAPERJ and  CNPq
is also gratefully acknowledged.

\end{document}